\documentstyle[11pt]{article}

\makeatletter
\@addtoreset{equation}{section}
\makeatother

\def\dalemb#1#2{{\vbox{\hrule height .#2pt
        \hbox{\vrule width.#2pt height#1pt \kern#1pt
                \vrule width.#2pt}
        \hrule height.#2pt}}}

\let\a=\alpha \let\b=\beta   \let\e=\epsilon

\def\nn{\nonumber} \def\bd{\begin{document}} \def\ed{\end{document}}
\def\ds{\documentstyle} \let\fr=\frac \let\bl=\bigl \let\br=\bigr
\let\Br=\Bigr \let\Bl=\Bigl 
\let\bm=\bibitem
\let\na=\nabla
\def\tU{{\widetilde U}}
\let\pa=\partial \let\ov=\overline
\def\ie{{\it i.e.\ }} 
\newcommand{\be}{\begin{equation}} 
\newcommand{\ee}{\end{equation}} 
\def\ba{\begin{array}}
\def\ea{\end{array}}
\def\ft#1#2{{\textstyle{{\scriptstyle #1}\over {\scriptstyle #2}}}}
\def\fft#1#2{{#1 \over #2}}
\def\del{\partial}
\def\R{{\bf R}}
\def\sst#1{{\scriptscriptstyle #1}}
\def\oneone{\rlap 1\mkern4mu{\rm l}}
\def\e7{E_{7(+7)}}
\def\td{\tilde}
\def\wtd{\widetilde}
\def\im{{\rm i}}
\def\bog{Bogomol'nyi\ }
\newcommand{\ho}[1]{$\, ^{#1}$}
\newcommand{\hoch}[1]{$\, ^{#1}$}
\newcommand{\bea}{\begin{eqnarray}} 
\newcommand{\eea}{\end{eqnarray}} 
\newcommand{\ra}{\rightarrow}
\newcommand{\lra}{\longrightarrow}
\newcommand{\Lra}{\Leftrightarrow}
\newcommand{\ap}{\alpha^\prime}
\newcommand{\bp}{\tilde \beta^\prime}
\newcommand{\tr}{{\rm tr} }
\newcommand{\Tr}{{\rm Tr} } 
\newcommand{\NP}{Nucl. Phys. }
\newcommand{\tamphys}{\it Center for Theoretical Physics,
Texas A\&M University, College Station, Texas 77843}
\newcommand{\ens}{\it Laboratoire de Physique Th\'eorique de l'\'Ecole
Normale Sup\'erieure\hoch{2}\\
24 Rue Lhomond - 75231 Paris CEDEX 05}

\newcommand{\auth}{I.V. Lavrinenko\hoch{\ddagger}, H.
L\"u\hoch{\dagger},  C.N. Pope\hoch{\ddagger1} and 
T.A. Tran\hoch{\ddagger}}

\begin{document}
\begin{flushright}
\hfill{CTP TAMU-9/97}\\
\hfill{LPTENS-97/05}\\
\hfill{hep-th/9702058}\\
\hfill{Feb. 1997}\\
\end{flushright}

\vspace{20pt}

\begin{center}
{\large {\bf Harmonic Superpositions of Non-extremal $p$-branes}} 

\vspace{30pt}
\auth

\vspace{15pt}

{\hoch{\dagger}\ens}

\vspace{10pt}
{\hoch{\ddagger}\tamphys}

\vspace{40pt}
\underline{ABSTRACT}
\end{center}

     The plot of allowed $p$ and $D$ values for $p$-brane solitons in 
$D$-dimensional supergravity is the same whether the solitons are extremal 
or non-extremal.  One of the useful tools for relating different points on 
the plot is vertical dimensional reduction, which is possible if periodic 
arrays of $p$-brane solitons can be constructed.  This is straightforward 
for extremal $p$-branes, since the no-force condition allows arbitrary 
multi-centre solutions to be constructed in terms of a general harmonic 
function on the transverse space.   This has also been shown to be possible 
in the special case of non-extremal black holes in $D=4$ arrayed along an 
axis.  In this paper, we extend previous results to include multi-scalar
black holes, and dyonic black holes.  We also consider their oxidation to
higher dimensions, and we discuss  general procedures for constructing the
solutions, and studying their symmetries.

{\vfill\leftline{}\vfill
\vskip	10pt
\footnoterule
{\footnotesize	\hoch{1} Research supported in part by DOE 
Grant DE-FG05-91-ER40633 \vskip	-12pt} \vskip 10pt
{\footnotesize
        \hoch{2} Unit\'e Propre du Centre National de la Recherche
Scientifique, associ\'ee \`a l'\'Ecole Normale Sup\'erieure et \`a
l'Universit\'e de Paris-Sud}} 

\pagebreak
\setcounter{page}{1}

\section{Introduction}

     M-theory and string theories admit a plethora of solitonic 
extended-object solutions.  These include both extremal $p$-branes, which 
satisfy a BPS saturation condition, and non-extremal, or ``black'' 
$p$-branes.  Various ways of enumerating and classifying the BPS-saturated 
states in toroidally-compactified theories have been developed.  One of 
these involves filling out a web of interconnections between the various 
points in a plot of $p$ versus spacetime dimension $D$, by making use of two 
different dimensional-reduction procedures.  The first of these, which is 
known as diagonal dimensional reduction, involves performing a Kaluza-Klein 
compactification of one or more of the spatial $p$-brane dimensions, thereby 
reducing $p$ and $D$ simultaneously \cite{dhis,lpss1}.  The second
procedure, known as  vertical dimensional reduction, instead involves
performing the Kaluza-Klein reduction on one or more of the dimensions in
the transverse space \cite{k,ghl,lps}.  This  lowers $D$, while keeping $p$
fixed.  In this latter reduction scheme, it is  necessary first to
construct a suitable configuration of $p$-branes in the  higher dimension
that is independent of the transverse coordinates for which  the
compactification is to be performed.  For extremal $p$-branes this is 
quite straightforward since arbitrary multi-centre solutions can be constructed,
owing  to the existence of a no-force condition between individual extremal 
$p$-branes \cite{cg}.  Thus one can make configurations in the higher
dimension in  which the centres are distributed uniformly over a line,
plane or hyperplane,  which can then be compactified by the usual
Kaluza-Klein procedure.

     It is of interest to try to achieve a similar web of interconnections 
for the case of non-extremal $p$-brane solitons.  It has been shown that 
there is a universal prescription for ``blackening'' any extremal $p$-brane 
to give a non-extremal one \cite{host,dlp}, and so the plot of the allowed
values of $p$ versus $D$ for such 
solutions will be the same as in the extremal case.  What at first sight is 
lacking, however, is a fully analogous set of reduction procedures for 
spanning the two-dimensional $(p,D)$ plane.  Actually there is no difficulty 
in achieving the trajectories of the diagonal dimensional reduction, since 
no features specific to extremal solutions were used in this case.  However, 
for the vertical reduction it was important that one should be able to 
construct the necessary multi-centre solutions in the higher dimension, and 
it is commonly held that the zero-force condition implied by extremality 
is essential for this.  In fact, as was argued in \cite{m,lpx}, this is not 
really the case.  The point here is that the ability to construct 
equilibrium multi-centre solutions with arbitrary positions for the centres
is an unnecessary luxury, if all one wants to do is to build $p$-brane
configurations that are uniformly distributed over a line, plane or
hyperplane.  It is in fact sufficient to be able to construct multi-centre
solutions where the centres are aligned in an infinite periodic array, along
the line, plane or hyperplane.  In any such configuration, there will always
be a net balance of forces on each individual $p$-brane, and so the system
will be in equilibrium, albeit an unstable one. However, this instability
need not be an obstacle to the explicit construction of the periodic
solutions.  In fact, by imposing the periodicity on the compactifying
coordinates, even the instability is eliminated. In \cite{lpx}, such
multi-centre solutions for dilatonic black holes in $D=4$ were obtained,
generalising previous results for black holes in Einstein gravity \cite{ik}
and the Maxwell-Einstein system \cite{g}.  A generalisation to periodic
multi-centre  $p$-branes in $(p+4)$ dimensions was also given in \cite{lpx};
these solutions  can be obtained by performing a diagonal oxidation of the
multi-centre black  holes in $D=4$.  Although the physical argument given
above indicates that periodic multi-centre generalisations should exist
for {\it  any} non-extremal $p$-brane, the complexity of the equations in
cases where $p < D-4$ seems to be too great to allow explicit solutions to
be found.

     In this paper, we shall investigate some broader classes of 
multi-centre non-extremal $p$-branes, although again restricted to the 
$D=p+4$ trajectory.  We begin by studying multi-scalar 
non-extremal black holes in $D=4$.  These correspond to configurations that 
are supported by a number of independently-specifiable charges carried by 
different 2-form field strengths, reducing to single-scalar black holes if 
the charges are set equal.  Then, we generalise these solutions to 
multi-scalar black $p$-branes in $(p+4)$ dimensions.  Another case that 
would be of great interest to study is the dyonic black hole in four 
dimensions \cite{gk}.  This is unusual in that, viewed as a bound state of
its two constituents, one with electric and the other with magnetic charge,
it has negative  binding energy \cite{gk} even when extremal.  In this
paper we are able to  obtain some special multi-centre solutions for the
dyonic black holes, but unfortunately  the general multi-centre solutions
with separated electric and magnetic charges seem to be too complicated to
construct  explicitly.  We do, however, present some general discussion on
approaches  to solving the equation, and also of solution-generating
symmetry groups.

\section{Multi-charge multi-centre black holes}

    In this section, we shall consider black-hole solutions in 
four-dimensional supergravity.  Thus the relevant part of the 
four-dimensional Lagrangian will be of the form
\be
{\cal L} = e R -\ft12 e (\del\vec\phi)^2 -\ft14 e\, \sum_{\a=1}^N 
e^{\vec a_\a \cdot \vec \phi}\, {F_2^{(\a)}}^2 \ ,\label{d4lag}
\ee
where $\vec\phi$ denotes the set of dilatonic scalar fields, $F_2^{(\a)}$ are 
$N$ 2-form field strengths, and $\vec a_\a$ are a set of constant vectors 
describing the couplings of the dilatonic scalars to the field strengths.
In the context of toroidal compactifications of $M$ theory, the expressions 
for $\vec a_\a$ for all the field strengths may be found in \cite{lpsol}.  
However, we should emphasise that our discussion of multi-centre solutions 
is more general, and need not necessarily be related to any supergravity
theory.  We shall be looking at solutions where the black holes carry
independent charges for the various field strengths $F^{(\a)}$, implying
that a corresponding number of scalar fields are non-vanishing. 

     Since we are looking for solutions describing an array of black holes 
aligned along an axis, we may assume the following axially-symmetric form 
for the metric ansatz:
\be
ds^2 = - e^{2U}\, dt^2 + e^{2K-2U}\, (dr^2 + dz^2) + e^{-2U}\, r^2\, 
d\theta^2\ ,\label{metric}
\ee
where $U$ and $K$ are both functions of $r$ and $z$ only.  (Actually, the 
most general axially-symmetric ansatz, after simplification by choosing 
coordinates appropriately, would have an independent function $e^{2B}$ 
rather than $e^{-2U}$ in the final term.  However, it can be shown that 
after making use of the equations of motion, one may choose coordinates such 
that $B=-U$ \cite{syn}.)  In this section, we shall consider the case where
all the  field strengths carry electric charges, and thus their potentials
may be  taken to be of the purely electric form
\be
A^{(\a)}= \gamma_\a\, dt\ ,\label{pot}
\ee
where the functions $\gamma_\a$ again depend on $r$ and $z$ only.  The dual
configurations where the field strengths carry purely magnetic charges 
instead are, as usual, very similar in structure.  The final solutions for
the metrics will be identical, and the solutions for the dilatonic scalars
will differ only by a sign.

     The multi-scalar solutions that are of interest to us here may be found 
by using the techniques described in \cite{lpmulti}.  For a set of field
strengths and scalars with couplings described by a generic set of vectors
$\vec a_\a$ in (\ref{d4lag}), the equations of motion cannot be decoupled,
and as far as we know they would not be solvable explicitly.  However, in
the supergravity theories obtained by the toroidal reduction of M-theory,
there are sets of vectors which, in the special case of 2-form field
strengths in $D=4$, satisfy the relations \cite{lpsol}
\be
\vec a_\a \cdot \vec a_\b = 4\delta_{\a\b} - 1\ .\label{adot}
\ee
This is precisely the condition that allows the system of equations of 
motion following from (\ref{d4lag}) to be diagonalised.  

     To solve the equations, it is convenient to define $\varphi_\a\equiv \vec 
a_\a \cdot \vec\phi_\a$, and to introduce new variables $\Phi_\a$ and $Y$ 
in place of $\varphi_\a$ and $U$; 
\be
\Phi_\a= -\varphi_\a + 2U\ ,\qquad Y= 2U -\ft14\, 
\sum_{\a=1}^N \Phi_\a\ ,\label{redefs}
\ee
where we are considering the case in which $N$ field strengths $F^{(\a)}$ 
carry charges.   It is straightforward to show that the equations of motion 
following from (\ref{d4lag}), with the ans\"atze (\ref{metric}) and 
(\ref{pot}), become
\bea
&&\nabla^2 Y=0\ , \qquad \nabla^2 \Phi_\a = 2\, e^{\Phi_\a}\, 
(\nabla\gamma_\a)^2\ ,\nn\\
&&\nabla^2 \gamma_\a = \nabla\Phi_\a\cdot\nabla\gamma_\a 
\ .\label{mseom}
\eea
Note that the Laplacian operator here is that of flat 3-dimensional space in 
cylindrical polar coordinates, acting on axially-symmetric functions, namely
\be
\nabla^2 =\fft{d^2}{dr^2} + \fft1{r}\, \fft{d}{dr} + \fft{d^2}{dz^2} \ ,
\ee
and $\nabla f \cdot \nabla g= f'\, g' + \dot f \, \dot g$, where a prime 
denotes a derivative with respect to $r$, and a dot denotes a derivative 
with respect to $z$.

    Later, in section 5, we shall discuss rather general procedures for
solving the equations. For now, we shall simply present the solutions that
are relevant to our discussion.  As in the previously studied cases of
axially-symmetric solutions to the Einstein \cite{syn,ik}, Einstein-Maxwell
\cite{g} and Einstein-Maxwell-Dilaton \cite{lpx} systems, we find here that
a broad class of solutions can be obtained in terms of axially-symmetric
harmonic functions in the $(r,\theta,z)$ space. Specifically, we find that
the equations of motion are satisfied if 
\bea
\nabla^2 Y&=&0\ ,\nn\\
e^{-\ft12\Phi_\a} &=&  \Big(e^{-\tU_\a} - c_\a^2 e^{\tU_\a}\Big)
\nn\\
\gamma_\a&=& c_\a\, e^{2\tU_\a}\, \Big(1-c_\a^2\, e^{2\tU_\a}\Big)^{-1}\ ,
\label{eom}\\
K'&=& -\fft{r}{4-N}\,(\dot Y^2 -{Y'}^2) -\ft14 r\, \sum_\a(\dot\tU_\a^2
-{\tU_\a}'^2) \ ,\nn\\
\dot K &=& \fft{2r}{4-N}\, \dot Y\, Y' +\ft12\, r\, \sum_\a \dot\tU_\a \,
\tU_\a' \ ,\nn
\eea
where the $c_\a$ are constants.  (Note that if $N=4$ the 
matrix of dot products (\ref{adot}) has vanishing determinant, implying that 
$Y$ vanishes.)  The functions $\tU_\a$ are arbitrary harmonic functions,
satisfying 
\be 
\nabla^2 \tU_\a=0\ .
\ee
The final two equations in (\ref{eom}) are automatically consistent with the 
solutions for the other functions, in the sense that the $z$ derivative of 
the former is equal to the $r$ derivative of the latter, by virtue of the 
harmonicity of $Y$ and $\tU_\a$.  Thus the solution for $K$ is reduced to 
quadratures, and the entire solution is given in terms of the $N+1$ 
independent arbitrary harmonic functions $Y$ and $\tU_\a$.  The parameters 
$c_\a$ characterise the charges carried by each of the field strengths 
$F_2^{(\a)}$, with $c_\a=0$ corresponding to a zero charge. We shall explain 
this in more detail below.

     The results presented above describe rather broad classes of 
axially-symmetric solutions to the equations of motion following from 
(\ref{d4lag}).  A subset of these will correspond to the configurations that 
we are seeking to construct, namely sets of multi-charge black holes arrayed 
along the $z$ axis.  In order to identify which are the solutions that 
describe these configurations, we need first to be able to recognise the basic 
single-centre non-extremal multi-charged black hole in this coordinate 
system.  In the previous non-scalar or single-scalar cases, it turned out 
that the required harmonic function corresponding to the single black hole 
configuration was precisely that describing the Newtonian gravitational 
potential of a uniform thin rod of mass $k$ and length $\ft12 k$
\cite{ik,g,lpx}. We find in this case that it is again this same harmonic
function that  describes the single-centre multi-charge black hole. 
Specifically, we should take
\be
\tU_\a =\tU\equiv \ft12 \log\fft{\sigma +\tilde\sigma -
k}{\sigma +
\tilde\sigma +k}\ ,\qquad Y=\ft12(4-N) \, \tU\ ,
\ee
where $\sigma=\sqrt{r^2 + (z-k/2)^2}$ and $\tilde\sigma=\sqrt{r^2 
+(z+k/2)^2}$, implying that the function $U$ appearing in the metric 
(\ref{metric}) is given by
\be
e^{2U}=e^{2\tU}\, \prod_\a\Big( 1-c_\a^2\, e^{2\tU}\Big)^{-1/2}\ .
\ee
(Note that indeed the function $Y$ vanishes in the degenerate $N=4$ case 
mentioned above, for which the determinant of the matrix in (\ref{adot}) 
vanishes.)
After performing the necessary integrals, one then finds that $K$ is given 
by
\be
K= \ft12 \log\fft{(\sigma+\tilde \sigma -k)(\sigma+\tilde \sigma 
+k)}{4\sigma\, \tilde\sigma}\ .
\ee
To see that this is describing the right solution, we note that in isotropic 
coordinates the standard single-centre multi-charge black hole is given by
\bea
ds^2 &=& -\fft{(1-\fft{\hat k}{R})^2}{(1+\fft{\hat k}{R})^2} \,
\prod_\a\Big(1+ \fft{4\hat k R}{(R+\hat k)^2}\, \sinh^2\mu_\a \Big)^{-1/2}
\, d\hat t^2 \nn\\
&&+\fft1{16} \Big(1+\fft{\hat k}{R}\Big)^4 \, \prod_\a \Big( 1 + 
\fft{4\hat k R}{(R+ \hat k)^2} \, \sinh^2\mu_\a\Big)^{1/2}\,
(d\rho^2 + dy^2 + \rho^2 \, d\theta^2)\ ,\label{iso}
\eea
where $R^2=\rho^2 + y^2$.  This metric can be recast into the form of the 
ansatz (\ref{metric}), by performing an holomorphic transformation from the 
complex coordinate $\xi=\rho+ \im\, y$ to the complex coordinate 
$\eta= r+ \im \, z$, in a manner analogous to that observed in the previous
cases \cite{lpx}. Guided by a comparison of the terms proportional to
$d\theta^2$, it is not hard to see that the required holomorphic
transformation is given by 
\be
\eta = \ft14 \Big(\prod_\a \cosh\mu_\a\Big)^{1/2}\, \Big(\xi 
-\fft{\hat k^2}{\xi} \Big)\ ,\label{holo}
\ee
where $\hat k= k\, (\prod_\a\cosh\mu_\a)^{-1/2}$, and $c_\a=\tanh\mu_\a$.  
Defining also a rescaled time coordinate $\hat t=t\, 
(\prod_\a\cosh\mu_\a)^{1/2}$, we find that the metric (\ref{metric})
assumes the form (\ref{iso}).  This has mass $m$ and charges $\lambda_\a$ 
given by
\be
m= \hat k \sum_\a \sinh^2\mu_\a + 2 \hat k N\ ,\qquad 
\lambda_\a= \ft12 \hat k \sinh2\mu_\a\ .
\ee
It is evident from this that extremality is achieved by sending $\hat k$ to 
zero, with one or more of the parameters $\mu_\a$ going to infinity
\cite{dlp}, in such a way that the mass and charges remain finite. Thus in
terms of the original parameters $k$ and $c_\a$ in the axially-symmetric
multi-scalar solutions, extremality is achieved if one or more of the $c_\a$
is equal to 1. 

     Having identified the single centre multi-scalar black hole solution
in the axially symmetric coordinate system that we are using here, it is
now straightforward to see how to generalise it to a multi-centre
solution.  Clearly, since the solutions are given in terms of arbitrary
harmonic functions, we may simply take a linear superposition of the
basic single-centre solutions described above, located at different points
$z_n$ along the $z$ axis.  For the time being, we shall continue to take
all the harmonic functions $\tU_\a$ to be equal, and
\be
\tU_\a = \tU\equiv \sum_n\ft12 \log\fft{\sigma_n +\tilde\sigma_n -
k_n}{\sigma_n +
\tilde\sigma_n +k_n}\ ,\qquad Y=\ft12(4-N)\, \tU \ ,\label{multicenu}
\ee
where $\sigma_n=\sqrt{r^2 + (z-z_n-k_n/2)^2}$ and $\tilde\sigma=\sqrt{r^2 
+(z-z_n+k_n/2)^2}$.  The solution for $K$ can now be shown to be 
\bea
K &=& \ft14 \sum_{m,n=1}^N \log\fft{[\sigma_m \td\sigma_n +
(z-z_m -\ft12 k_m)(z-z_n+ \ft12 k_n) + r^2]}{[\sigma_m \sigma_n +
(z-z_m -\ft12 k_m)(z-z_n- \ft12 k_n) + r^2]}\label{multisol}\\
&&+\ft14 \sum_{m,n=1}^N \log\fft{[\td\sigma_m \sigma_n +
(z-z_m +\ft12 k_m)(z-z_n- \ft12 k_n) + r^2]}{[\td\sigma_m \td\sigma_n +
(z-z_m +\ft12 k_m)(z-z_n+ \ft12 k_n) + r^2]}\ ,\nonumber
\eea
where $\sigma_n^2 = r^2 + (z-z_n-\ft12 k_n)^2$ and $\td \sigma_n^2 =
r^2 +(z-z_n+\ft12 k_n)^2$. 
In the vicinity of each of the points $z_n$, the solution is equivalent to
the single-centre multi-scalar black hole described previously, and thus
the configuration characterised by (\ref{multisol}) describes a
multi-centre multi-scalar black hole solution, with an independent
multi-scalar black hole at each of the locations $z_n$.  

     The global structure of such non-extremal multi-black hole solutions
was discussed in the simpler cases of Einstein and Einstein-Maxwell
solutions in \cite{g2,hs}.  As one might expect, if a solution with a finite
number of centres is considered, or indeed any configuration that is not
of a simple periodic nature, then the solution suffers from
singularities.  This is because in such cases there is no balance of
forces on each of the black holes, and thus the system can remain in
static equilibrium only if there are ``struts'' linking the black holes,
which supply the necessary additional forces necessary to hold the system
in place.  These struts manifest themselves in the form of conical
curvature singularities on the $z$ axis in the gaps between the black
holes \cite{g2,hs}.  It is clear, however, that if the black holes are
arrayed in a simple periodic fashion along the $z$ axis, the net force on
each black hole will be zero, and thus no struts are necessary to maintain
the static equilibrium.  Indeed, under precisely this circumstance, the
conical singularities disappear.  In a similar manner, we find in the
present case of the multi-scalar non-extremal black hole solutions that
the metric is free of conical curvature singularities if the ``Newtonian
masses'' $k_n$ are chosen equal, the locations $z_n$ are taken to be of
the periodic form $z_n= z_0+ n\, b$, and the index $n$ is allowed to
range over the entire set of integers, positive, negative, and zero. 
If the spacing $b$ is chosen to be sufficiently small, then seen at a
large distance from the $z$ axis, the solution will then limit to a
single-centre multi-scalar black hole in the remaining three directions
orthogonal to $z$.  

     The situation described above may be viewed as a vertical
dimensional reduction of multi-scalar non-extremal black holes from four to
three dimensions.  In the continuum limit where $b$ is very small, one can
show \cite{m,lpx} that the expressions (\ref{multicenu}) for $\tU_\a$
become the $z$-independent function $\beta\, \log r$, where $\beta=k/b$
and $b$ is the inter-black-hole spacing.  Substitution into the equations
for $K$ in (\ref{eom}) then implies that $K=\beta^2\, \log r$.  From
(\ref{redefs}) we can solve for $U$, giving
\be
e^{2U}= r^{2\beta} \, \prod_\a (1 - c_\a^2\, r^{2\beta})^{-1/2}\ .
\ee
Thus in this continuum limit, the four-dimensional metric $ds_4^2$ given by 
(\ref{metric}) becomes independent of $z$, allowing us to perform a
Kaluza-Klein reduction to $D=3$, with 
\be
ds_4^2 = e^{\varphi}\, ds_3^2 +
e^{-\varphi}\, dz^2\ ,\label{d43}
\ee
where $\varphi$ here denotes the Kaluza-Klein scalar 
that comes from the dimensional reduction of the 4-dimensional metric.  Note 
that here, and in all other discussions of dimensional reduction in this 
paper, we are working always with the Einstein-frame forms for the metrics.
The dimensionally-reduced multi-charge black hole metric in $D=3$ is
therefore given by 
\be 
ds_3^2 = -r^{2\beta^2}\, dt^2 + r^{2\beta^2 -4\beta}\, \prod_\a (1 
-c_\a^2\, r^{2\beta} )\, (r^{2\beta^2}\, dr^2 + r^2\, d\theta^2)\ .
\label{d3metric}
\ee
This is, in general, not quite the ``usual'' kind of $N$-charge black hole 
solution in $D=3$.  The reason for this is that usually a solution with $N$ 
independent charges would involve the excitation of $N$ independent 
dilatonic scalars.  This is indeed the case in our original $D=4$ solutions 
(except in the degenerate case $N=4$, where only three independent scalars 
are excited.)  However upon reduction to $D=3$ a new scalar, namely the 
Kaluza-Klein scalar $\varphi$ above, is also excited.  The condition for a 
normal $N$-charge, $N$-scalar solution is that all projections of the full 
set of $(11-D)$ dilatonic scalars orthogonal to the $N$ dilaton vectors 
$\vec a_\a$ in $D$ dimensions should be zero.  In the present context, it is 
not hard to see that this would imply that we should have $\sum_\a
\varphi_\a - (N-4) \varphi=0$.  Comparing (\ref{metric}) with (\ref{d43}),
we see that $\varphi=2 U-2K$, and thus it follows, after substituting in our
solutions for $U$, $Y$ and $K$ in the continuum limit, that the condition
for the additional scalar degree of freedom to be unexcited in $D=3$ is that
$\beta=1$.  This implies that the separation $b$ between the black holes is
equal to the length $k$ of the Newtonian rods.  In fact the Newtonian
potential now describes the gravitational field of a single rod of infinite
length, and mass density $\ft12$. Not surprisingly, the metric in this case is
rather pathological, and in fact describes a continuous circle of $D=3$
black holes in the extremal limit, rather than a single black hole.  (This
was discussed in detail for the single-charge case in \cite{lpx}.)  Thus it
would seem that one should really view the more generic solutions with
$\beta\ne1$, and hence with one more independent scalar excitation than 
normal, as the more appropriate generalisations of black-hole solitons
to $D=3$. In particular, we should consider the case where $\beta<1$, so
that the separation between the black holes in $D=4$ is less than their
horizon radius.

     Note that more general multi-centre solutions are also possible.
Since the most general solutions that we have constructed allow independent
harmonic functions for each kind of scalar field, we may allow the
various different kinds of charge to reside at different sets of locations
along the $z$ axis.  By choosing periodic arrays of centres appropriately,
we may thus obtain three-dimensional multi-charge solutions
by vertical reduction from $D=4$. 

\section{Multi-charge multi-centre black $(D-4)$-branes}

     As we have remarked in the introduction, the existence of special
classes of static multi-centre non-extremal $p$-branes, namely those with a
simple periodicity along an axis, plane or hyperplane, is guaranteed on
general grounds since there will be a net balance of forces on each
individual $p$-brane.  What is not guaranteed, however, is that there should
exist such multi-centre solutions for arbitrarily located sets of non-extremal
$p$-branes.  Nevertheless, in the case of black holes in four dimensions,
it turns out that axially-symmetric solutions with black holes disposed
arbitrarily along an axis do exist.  The associated imbalance of forces in
general is counterbalanced by the occurrence of ``struts'' that are
associated with conical-type delta-function curvature singularities along
the axis between the black holes.  In a case where the transverse space has
dimension greater than 3, for example black holes in five dimensions, it is
not at all clear that analogous axially-symmetric solutions for arbitrarily
disposed sets of centres will exist; in particular, the notion of conical
singularities with curvature zero except on submanifolds no longer exists. 
Thus it is quite likely that a description of multi-centre non-extremal
$p$-branes in terms of arbitrary axially-symmetric harmonic functions is no
longer possible when the transverse space has dimension greater than 3, 
although there should certainly exist simply-periodic solutions.  The
equations of motion for such systems were given in \cite{lpx}, and
preliminary investigations indeed seemed to confirm that finding solutions
in terms of harmonic functions would be problematical.  

     If, therefore, we restrict attention to cases where the transverse
space has dimension 3, then this implies that we should consider
$(D-4)$-branes in $D$ dimensions.  Black holes in $D=4$ are thus the
simplest case.  The form of the axially-symmetric metrics for $(D-4)$-branes
is like (\ref{metric}), but with the addition of $p$ spatial world-volume
dimensions: 
\be
ds^2 = -e^{2U} \, dt^2 + e^{2A}\, dx^i\, dx^i + e^{2K-2U}\, (dr^2 + dz^2) +
e^{2B}\, r^2\, d\theta^2 \ ,\label{diagmetric}
\ee
where $U$, $K$, $A$ and $B$ are all functions of $r$ and $z$.  The curvature for
such metrics was calculated in \cite{lpx}, and from this it is a
straightforward matter to obtain the equations of motion for the system
with multiple scalars and field strengths described by (\ref{d4lag}), now
taken to be in $D$ dimensions.  However in practice, a simpler way to
construct the required solutions in $D$ dimensions is to make use of the
fact that they can be dimensionally reduced to $D=4$ by compactifying the
additional $p$ world-volume dimensions, when they will become black-hole
solutions of the kind we have already constructed in section 2.  Thus we may
obtain the multi-charge multi-centre non-extremal $p$-branes by inverting
the process of Kaluza-Klein dimensional reduction.  This oxidation method
was used in \cite{lpx} in order to construct such solutions in the
single-charge case. 

     In order to discuss the oxidation of the four-dimensional black hole
solutions, it is useful to consider the step in the Kaluza-Klein reduction
process in which we pass from $D$ dimensions to $(D-1)$ dimensions.  In $D$
dimensions, the dilaton vectors $\hat{\vec a_\a}$ associated with the $N$
field strengths must satisfy 
\be
\hat{\vec a_\a} \cdot \hat{\vec a_\b} = 4\delta_{\a\b} - \fft{2(D-3)}{D-2}
\label{vecid}
\ee
in order that the equations of motion factorise into a solvable form
\cite{lpmulti}.  Upon reduction to $(D-1)$ dimensions, the dilatons
$\hat{\vec\phi}$ will be augmented by the scalar $\varphi_{\sst D-1}$ coming
from the Kaluza-Klein reduction of the metric, and we may represent the full 
set by the vector $\vec\phi=(\hat{\vec\phi},\varphi_{\sst D-1})$.  The
Kaluza-Klein reduction of the kinetic term for the $\a$'th field strength
will yield an additional coupling of the new scalar field, implying that in
$(D-1)$ dimensions the dilaton vector is given by $\vec a_\a = (\hat{\vec
a_\a}, 2\a_{\sst D -1})$, where $\a_{\sst D-1} \equiv
(2(D-2)(D-3))^{-1/2}$.  It follows that the quantities $\varphi_\a=\vec
a_\a\cdot \vec\phi$ and $\hat\varphi_\a=\hat{\vec a_\a}\cdot \hat{\vec\phi}$
in $(D-1)$ and $D$ dimensions, analogous to the ones we defined in $D=4$ in
section 2, are related by $\hat\varphi_\a =\varphi_\a - 2\a_{\sst D-1}\,
\varphi_{\sst D-1}$.  Since we are wanting to describe the
higher-dimensional solutions that arise from oxidation of the
lower-dimensional ones, it must also be the case that the additional scalar
field that we acquire in the step from $D$ to $(D-1)$ dimensions must not 
lead to any additional independent scalar field in the lower dimensional
solution.  Thus the linear combination of the new scalar $\varphi_{\sst 
D-1}$ and the old scalars $\hat{\vec\phi}$ 
that is orthogonal to those combinations that are excited in the
lower-dimensional solution must vanish.  This is the combination that is
orthogonal to all the dilaton vectors $\vec a_\a$ in $(D-1)$ dimensions, \ie
the combination parallel to the vector 
\be
\vec n = \Big(\sum_\a \hat{\vec\a_\a}, \fft{1}{\a_{\sst
D-1}}\, (\fft{N(D-3)}{D-2} -2)\Big)\ .
\ee
This implies that the scalars in $(D-1)$ dimensions must satisfy 
\be
\sum_\a \varphi_\a +\fft{1}{\a_{\sst D-1}} \, \Big(\fft{N(D-4)}{D-3} -2\Big)
\varphi_{\sst D-1} = 0\ .
\ee

     With the above conditions satisfied, it is now a straightforward
matter to implement the oxidation procedure recursively, using the standard
relation
\be
ds_{\sst D}^2 = e^{2\a_{\sst D-1}\varphi_{\sst D-1}}\, ds_{\sst D-1}^2 +
e^{-2(D-3)\a_{\sst D-1}\varphi_{\sst D-1}}\, dz^2 \label{kkred}
\ee
between the $D$-dimensional and $(D-1)$-dimensional metrics.  This enables
us to express the multi-charge multi-centre black $(D-4)$-brane solutions in
$D$ dimensions in terms of the quantities $U$, $K$ and $\varphi_\a$ given
in section 2 for the four-dimensional black hole solutions.  In terms of
these, we find the following $D$-dimensional metric:
\be
ds_{\sst D}^2 = -e^{2U+(D-4) V}\, dt^2 + e^{-2V}\, \sum_{i=1}^p dx^i\, dx^i
+ e^{-2U+(D-4) V}\, \Big( e^{2K}(dr^2+dz^2) + r^2\, d\theta^2) \Big)\ ,
\label{dmetric}
\ee
where 
\be
V=\fft{1}{(D-2)(4-N)} \, \sum_\a \varphi_\a\ .\label{vdef}
\ee
This enables us to describe axially-symmetric multi-centre configurations of 
non-extremal multi-charge $(D-4)$-branes in $D$ dimensions.

\section{Oxidation to $D=11$}

      Having obtained 4-dimensional multi-charge multi-centre non-extremal
black holes from the bosonic Lagrangian (\ref{d4lag}) in section 2, we
now study how these solutions are embedded in the 4-dimensional
maximal supergravity theory.  Since the maximal supergravity in $D=4$
is nothing but a consistent Kaluza-Klein dimensional reduction of
11-dimensional supergravity on the 7-torus, we may oxidise these
solutions back to $D=11$, thus giving an 11-dimensional interpretation of
the solutions that we have obtained.

       There are a total of twenty-eight 2-form field strengths in
4-dimensional maximal supergravity, each of which can be used to construct
either an electrically-charged or a magnetically-charged black hole.  These
56 black hole solutions form \cite{ht} a multiplet under the $E_7$ group of
the supergravity theory \cite{cj,c}.  To be more precise, they form a
56-dimensional multiplet under the Weyl group of $E_7$ \cite{lpsweyl}.  As we 
mentioned in section 2, when the dilaton vectors for the 2-form field
strengths satisfy the dot products (\ref{adot}), there can exist a
consistent truncation of the bosonic Lagrangian to (\ref{d4lag}). (It
should be remembered that not all sets of field strengths that have this
property will lead to a consistent truncation, but at least one
member of their Weyl multiplet will \cite{lpsweyl}.)  The maximal number of
participating 2-form field strengths in the Lagrangian (\ref{d4lag}) in
$D=4$ that can be obtained by consistent truncation of the maximal supergravity
theory is $N_{\rm max}=4$.  As was shown in \cite{lpsweyl}, the dimension of
the duality multiplet of this $N=4$ solution is 630.  In other words,
there are 630 choices of sets of four 2-form field strengths whose dilaton 
vectors satisfy the dot products (\ref{adot}).  In this
paper, we shall consider just one example.  We are particularly
interested in the case where the 2-form field strengths come directly
from the dimensional reduction of the 4-form field strength in $D=11$,
rather than those coming from the metric.  For these $p$-brane
solutions, the oxidation procedure will give rise to $p$-branes or
intersecting $p$-branes in all higher dimensions.  By contrast, for the
solutions involving field strengths coming from the metric, the
oxidation would give rise to boosts or topological twists in the
metric in higher dimensions.

        We follow the notation of \cite{lpsol} for dimensional reduction of
11-dimensional supergravity on the 7-torus, which was performed by the
iteration of a one-step reduction procedure.  The 4-form field strength
gives rise to seven 3-form field strengths $F_3^{(i)}$, which are dual to
1-forms; twenty-one 2-forms $F_2^{(ij)}$ and thirty-five 1-forms
$F_1^{(ijk)}$. The dilaton vectors for these field strengths are denoted by
$\vec a_{i}$, $\vec a_{ij}$ and $\vec a_{ijk}$ respectively, where the
indices $i, j$ and $k$ run over the seven internal compactified coordinates.
The explicit forms of these vectors can be 
found in \cite{lpsol}.  There are a further seven 2-forms ${\cal F}_2^{(i)}$
and twenty-one 1-forms ${\cal F}_1^{(ij)}$ ($i <j$) coming from the
metric.  In this paper, we shall consider a 4-charge black hole whose 
charges are carried by the field strengths $F_2^{(12)}$, $F_2^{(34)}$,
$*F_2^{(13)}$ and $*F_2^{(24)}$, and corresponding dilaton vectors
$\vec c_1 = \vec a_{12}$, $\vec c_2 = \vec a_{34}$, $\vec c_3 = -\vec
a_{13}$ and $\vec c_4 = -\vec a_{24}$.  From their expressions given in 
\cite{lpsol}, it is easy to verify that these dilaton vectors satisfy
(\ref{adot}), and also that the bosonic Lagrangian
can be consistently truncated to (\ref{d4lag}), with the standard
electric or magnetic ansatz for the field strengths.  If we consider
an electrically-charged  solution, the Hodge duals for the field strengths
$*F_2^{(13)}$ and $*F_2^{(24)}$ imply that in terms of the original
fields $F_2^{(13)}$ and $F_2^{(24)}$, these field strengths carry
magnetic charges; if we consider instead a magnetic solution, these field
strengths will describe electric charges.  It is this Hodge dualisation that 
is responsible for the dilaton vectors $\vec c_3$ and $\vec c_4$ having the 
minus signs that we indicated above.

       The 4-dimensional 4-charge multi-centre non-extremal black hole
solutions were obtained in section 2.  It is straightforward to
oxidise the solutions back to $D=11$, by applying (\ref{kkred}) iteratively.  
Defining $H_\a = e^{-\wtd U_\a}
-c_\a^2 e^{\wtd U_\a}$, we find that the metric of the corresponding
11-dimension solution is given by 
\bea
ds_{11}^2 &=& \Big(\fft{H_3H_4}{H_1H_2}\Big)^{\ft16}\, ds_4^2 +
     \Big(\fft{H_2 H_3^2}{H_1^2 H_4} \Big)^{\ft13} dz_1^2 +
     \Big(\fft{H_2 H_4^2}{H_1^2 H_3} \Big)^{\ft13} dz_2^2 +
     \Big(\fft{H_1 H_3^2}{H_2^2 H_4} \Big)^{\ft13} dz_3^2\nonumber\\
&&  + \Big(\fft{H_1 H_4^2}{H_2^2 H_3} \Big)^{\ft13} dz_4^2 +
     \Big(\fft{H_1 H_2^2}{H_3^2 H_4} \Big)^{\ft13} (dz_5^2
                    + dz_6^2 + dz_7^2)\ ,\label{d11}
\eea
where $ds_4^2$ is the 4-dimensional metric obtained in section 2.  Note
that if all $H_\a=1$ except for $H_1$, the coordinates $(z_1, z_2)$ become
the world-volume spatial coordinates of a membrane solution in $D=11$.  Since
there are translational symmetries of the remaining five compactifying
coordinates as well, it implies that the metric (\ref{d11}) then describes a
5-plane of uniformly distributed non-extremal membranes.  If instead only
$H_2 \ne 1$, then the metric describes a 5-plane of non-extremal
membranes with $(z_3, z_4)$ as the world-volume spatial coordinates.  The story
is slightly different with the other two cases.  If only $H_3 \ne 1$, then
the solution describes a 2-planes of non-extremal 5-branes with $(z_2, z_4,
z_5, z_6, z_7)$ as the world-volume spatial coordinates; if only $H_4 \ne 1$, it
describes a 2-plane of non-extremal 5-branes with 
$(z_1, z_3, z_5, z_6, z_7)$ as the world-volume spatial coordinates.  Thus 
in general the metric (\ref{d11}) describes the intersection of two membranes
and two 5-branes in eleven dimensions.   Oxidations of isotropic
non-extremal black holes were considered in \cite{ct}.  In this paper, our
solutions depend on a set of harmonic functions $\wtd U_\a$, which
generalises previous results.  In the extremal limit, the functions $H_\a$
themselves become harmonic functions on the 3-dimensional space $(r, z,
\theta)$, which enables one to interpret 4-dimensional single-centre
single-scalar or multi-scalar black hole solutions as bound states of
singly-charged black holes \cite{rahm}. 

         In the above discussion, we considered the electric
four-dimensional solutions (with the pre-dualisation of the 3'rd and 4'th
field strengths, of course).  If we instead consider magnetic
solutions, the 11-dimensional oxidation is still given by (\ref{d11}),
except this time we have $H_\a \rightarrow 1/H_\a$.  Thus in this
case, the $H_1$ and $H_2$ will instead be associated with 5-branes in
$D=11$, and $H_3$ and $H_4$ with membranes.   As we mentioned earlier, there
are more choices of sets of field strengths that give rise to a consistent
truncation of the bosonic Lagrangian to (\ref{d4lag}). For example, we could
instead use $F_2^{(12)}$, $F_2^{(34)}$, $F_2^{(56)}$ and $*{\cal F}_2^{(7)}$
\cite{lpsol}.  The four-dimensional metric is identical to the previous
solutions; however, since we are now using a different 
field-strength configuration, the oxidation to $D=11$ will be quite
different.  The involvement of the 2-form ${\cal F}_2^{(7)}$ coming from the
metric implies a twist or a boost in the 11-dimensional metric.  Note that
the 4-dimensional 4-charge black holes always involve at least one dualised
field strength.  In other words, in terms of the original fields, the
solutions always involve both electric and magnetic charges.  Such solutions
were called dyonic solutions of the first type in \cite{lpsol},
since each individual field strength nevertheless carries only one type of
charge.  In section 5.3, we shall consider non-extremal dyonic black hole
solutions of the second type, where a single 2-form field strength carries both
electric and magnetic charges. 

\section{Solutions and symmetries}

    In this section, we shall consider the general question of how to 
construct solutions to the systems of equations that arise for the 
axially-symmetric configurations.  To begin, we present a procedure for 
constructing a class of solutions to the single-scalar equations of motion, 
expressed in terms of two independent harmonic functions. In fact these 
solutions coincide, in the case $N=1$, with the general $N$-scalar solutions 
presented in the previous section.  Then we shall consider a different 
approach to obtaining these solutions, which exploits an $O(2,1)$ symmetry 
of the equations.  In a subsequent subsection, we shall then apply similar 
considerations to the more complicated case of single-scalar dyonic black 
hole solutions.

\subsection{Single-scalar solutions}

We begin by considering the simplest 
situation, discussed in \cite{lpx}, where there is just a single 2-form field
strength and a single scalar field.  The four-dimensional Lagrangian is then 
given by
\be
{\cal L} = e R -\ft12 e (\del\phi)^2 -\ft14 e\, e^{-a\phi} \, F^2\ .
\label{sslag}
\ee
Substituting the metric ansatz (\ref{metric}) and the ansatz $A=\gamma\, dt$ 
for an electric charge, we obtain the equations of motion
\bea
\nabla^2 U &=& \ft14 \, e^{-a\phi-2U} \, (\nabla\gamma)^2 \ ,\qquad
\nabla^2\phi =\ft12 a\, e^{-a\phi-2U} \, (\nabla\gamma)^2 \ ,\nn\\
\nabla^2 \gamma &=& \nabla\gamma\cdot \nabla(a \phi + 2U) \ ,\label{sseom}
\eea
together with the equations that determine $K$ in (\ref{metric}), namely
\bea
K'&=& \ft14 r\, e^{-a\phi-2U}\,(\dot\gamma^2-{\gamma'}^2) +r\, (U'^2 -
\dot U^2) +\ft14 r\, (\phi'^2 -\dot \phi^2)\ ,\nn\\
\dot K &=& -\ft12 r\, e^{-a\phi-2U}\, \dot\gamma\, \gamma' + 2 r\, \dot U\, 
U' +\ft12 r\, \dot\phi\, \phi' \ .\label{sskeom}
\eea
In order to solve the equations, we need only consider the system given in 
(\ref{sseom}), since the solution for $K$ then follows from (\ref{sskeom}) 
by quadratures.  In (\ref{sseom}), we do not yet need to assume that the 
functions $U$, $\phi$ and $\gamma$ are independent of the axial angle 
$\theta$, and so the Laplacian $\nabla^2$ and the gradient operator $\nabla$ 
itself can be thought of for now as the ordinary 3-dimensional operators in 
the flat transverse space.  We shall proceed by looking for ways to 
solve the equations (\ref{sseom}).  It is convenient first to introduce two 
new functions $f$ and $g$ in place of $U$ and $\phi$, defined by
\be
f= a \phi + 2U\ ,\qquad g = \phi -2a\, U\ .\label{fdef}
\ee
In terms of these, the equations of motion (\ref{sseom}) become
\be
\nabla^2 f =\ft12 \Delta\, e^{-f}\, (\nabla \gamma)^2\ ,\qquad
\nabla^2 g = 0 \ ,\qquad \nabla^2 \gamma = \nabla\gamma\cdot \nabla f
\ .\label{feom}
\ee

     One approach to solving (\ref{feom}) is to note that the 
solutions obtained in \cite{lpx} are such that $U$, $\phi$ and $\gamma$ are
all  expressed as certain algebraic functions of a single harmonic function 
$\tU$.  In fact, from the form of the equations (\ref{feom}), we see that $g$ 
is already an harmonic function, and furthermore that it does not appear at
all in the equations of motion for $f$ and $\gamma$.  Thus it is natural to
try a more general ansatz, in which we assume that $f$ and $\gamma$ are both
algebraic functions of a second, independent, harmonic function $\psi$, \ie
$f=f(\psi)$ and $\gamma=\gamma(\psi)$. Since $\nabla^2 \psi$ is assumed to
vanish, it follows that the equations of motion for $f$ and $\gamma$ in
(\ref{feom}) become 
\be
\fft{d^2 f}{d\psi^2} = \ft12 \Delta\, e^{-f} \Big(\fft{d\gamma}{d\psi}\Big)^2
\ ,\qquad \fft{d^2 \gamma}{d\psi^2} = \fft{df}{d\psi}\, \fft{d\gamma}{d\psi}
\ .\label{f1f2eom}
\ee
The latter can be integrated once to give $d\gamma/d\psi = 
(c/\sqrt\Delta) \, e^f$, where $c$ is a constant whose normalisation is 
chosen for later convenience.  Substituting this into the first equation in 
(\ref{f1f2eom}) gives the Liouville equation
\be
\fft{d^2 f}{d\psi^2}=  \ft12\, c^2 \, e^f \ .\label{feq}
\ee
This has the solution
\be
e^{-\ft12 f}= e^{-\psi} - c^2\, e^{\psi}\ ,\label{lsol}
\ee
where the constants of integration are absorbed into the freedom to scale 
and shift the arbitrary harmonic function $\psi$ by constants.  Using 
the Liouville 
equation (\ref{feq}) we can integrate the equation for $\gamma$ to
give $\gamma= 2/(\sqrt\Delta c)\, df/d\psi + \hbox{const}$.  Since 
$\gamma$ is a potential for the electric field we can choose the constant 
arbitrarily.  Making a convenient choice, we find that the solution is
\be
\gamma= \fft{2c}{\sqrt\Delta} e^{2\psi}\, (1- c^2 e^{2\psi})^{-1}\ .
\label{gamsol}
\ee
Thus the entire solution for $U$, $\phi$ and $\gamma$ is given in terms of 
two independent arbitrary harmonic functions $g$ and $\psi$.  The equations 
for $K$ in (\ref{sskeom}) reduce to
\bea
K' &=& \fft{r}{4\Delta} ( 4\psi'^2 -4\dot\psi^2 + g'^2 -\dot g^2)\ ,\nn\\
\dot K &=& \fft{r}{2\Delta} (\dot g\, g' + 4 \dot\psi \psi')\ .\label{keq}
\eea

     The previous solutions obtained in \cite{lpx} correspond to the case
where  the two harmonic functions $g$ and $\psi$ are chosen to be
proportional to a single harmonic function $\tU$, with $\psi=\tU$ and
$g=-2a \, \tU$. There are other special cases too, such as $\psi=0$ which
implies that 
$\gamma$ is simply a constant, and hence there is no electric charge.  Since 
$f$ is also constant in this case, it implies that $U$ and $\phi$ are both 
harmonic, and are proportional to the remaining harmonic function $g$.
Another special case is $g=0$, implying that $\phi$ and $U$ are related, 
$\phi=2 a U$, and given in terms of the remaining harmonic function $\psi$.
Finally, we note that if $c=0$ the electric charge again vanishes, and 
$\phi$ and $U$ become independent harmonic functions.  The special case when 
$\phi$ also vanishes corresponds to a solution of the pure Einstein 
equations.

\subsection{Solution-generating symmetry of the single-scalar system}

     As another approach to solving the equations of motion, it is useful to 
consider the symmetries of the system of equations (\ref{feom}).  
Since the equation for $g$ is independent of the others, we may focus 
principally on the equations of motion for $f$ and $\gamma$.  It is easy to 
see that these can be derived from the Lagrangian
\be
L = (\nabla f)^2 - \Delta\, e^{-f}\, (\nabla\gamma)^2 \ .\label{fgamlag}
\ee
Let us now introduce the
three fields $X$, $Y$ and $Z$, defined by $X+Y=2 e^{-f/2}$, $X-Y= 2 e^{f/2}
-\ft12 \Delta \gamma^2\, e^{-f/2}$ and $Z=\sqrt\Delta\, \gamma\,e^{-f/2}$.  
In terms of these, we have 
\bea
&&X^2-Y^2+Z^2=4\ ,\label{const}\\
&&L=-(\nabla X)^2 + (\nabla Y)^2 -(\nabla Z)^2\ .\label{so21lag}
\eea
The constraint (\ref{const}) and the Lagrangian 
(\ref{so21lag}) have a manifest $O(2,1)$ symmetry.  In terms of the original 
fields $f$ and $\gamma$, this symmetry indeed gives an invariance of the 
Lagrangian (\ref{fgamlag}), but owing to the choice of parameterisation
we should restrict to transformations that do not reverse the sign of 
$e^{f/2}$ ({\it i.e.}\ we should restrict to transformations where $X+Y$ is 
non-negative).  Thus the allowed symmetry transformations are of the form 
$O(2,1)/J$, where $J$ is the antipodal map $(X,Y,Z)\rightarrow 
(-X,-Y,-Z)$.   Since $J$ is central in $O(2,1)$, it follows that $O(2,1)/J$ 
is a group.  If an $O(2,1)$ transformation makes $X+Y$ negative, then 
$J$ is used to restore its positivity.\footnote{We are grateful to G.W.
Gibbons for discussions about the symmetry group of the Lagrangian
(\ref{fgamlag}).} 

     The symmetry group $O(2,1)/J$ has two disconnected components, one of
which is $SO(2,1)$ and the other is reached by performing the discrete 
``time-reversal'' transformation $Y\rightarrow -Y$.
Since $SL(2,R))$ is homomorphic to $SO(2,1)$, we can also express the
$SO(2,1)$ subgroup in a manner analogous to a fractional linear
transformation.  To see this, let us introduce the field $\chi=-\im \gamma$.
Temporarily treating $\chi$ as a real field, we may rewrite the Lagrangian
(\ref{fgamlag}) as 
\be
L=-\fft{ \nabla \tau \cdot \nabla \bar\tau}{(\tau-\bar\tau)^2}\ , 
\label{taulag}
\ee
where we have defined
\be
\tau = \ft12\sqrt\Delta\, \chi + \im \, e^{\ft12 f}\ .\label{taudef}
\ee
This is invariant under the fractional linear $SL(2,\R)$ transformations
\be
\tau\longrightarrow \fft{a \tau + b}{c \tau + d}\ ,\label{frac}
\ee
where $ad-bc=1$.  Rewriting this transformation in terms of its action on 
the fields $\gamma$ and $f$, which involves also sending $b\rightarrow -\im 
b$ and $c\rightarrow \im c$ in order to make the transformation real again, 
we find
\bea
\gamma &&\longrightarrow\quad \fft{2((\ft12 a\sqrt\Delta\, \gamma + b)(\ft12 c 
\sqrt\Delta\, \gamma + d) - a c\, e^f)}{\sqrt\Delta((\ft12 c 
\sqrt\Delta\, \gamma + d)^2 -c^2\, e^f)}\ ,\nn\\ 
e^{\fft12 f} &&\longrightarrow \quad\fft{e^{\fft12 f}}{(\ft12 c 
\sqrt\Delta\,\gamma + d)^2 -c^2\, e^f} \ ,\label{sl2rtrans}
\eea
where again $a d-bc =1$.

     The $O(2,1)/J$ symmetry provides a powerful way of constructing new 
solutions from old ones.  In particular, the ``time-reversal'' symmetry
$Y\rightarrow -Y$ of (\ref{const}) and (\ref{so21lag}) implies that
\be
 \gamma\longrightarrow -\fft{\gamma}{e^f - \ft14 \Delta \, \gamma^2} \ ,\qquad
e^{\ft12 f}\longrightarrow \fft{e^{\ft12 f}}{e^f -\ft14\Delta\, \gamma^2}
\ .\label{inv}
\ee
(Note that $Y\rightarrow -Y$ is contained in $O(2,1)/J$, but is not in the 
subgroup $SO(2,1)$ that is connected to the identity.)
Let us apply this to the simple solution
\be
\gamma= -\fft{2}{\sqrt\Delta}\, c\ ,\qquad f=-2\psi\ ,\label{uncharged}
\ee
where $c$ is a constant and $\psi$ is an harmonic function.  It is manifest 
that (\ref{uncharged}) satisfies the equations of motion given in 
(\ref{feom}).  In fact, it describes an uncharged solution, \ie a solution
of the pure Einstein-Dilaton equations.  Under the transformation 
(\ref{inv}), it becomes precisely the charged Einstein-Maxwell-Dilaton 
solution given in (\ref{gamsol}) and (\ref{lsol}).  Thus the discrete 
$Y\rightarrow -Y$ transformation in the $O(2,1)/J$ 
symmetry of the equations of motion maps between uncharged and charged 
solutions.  We note also that the expressions for 
$K'$ and $\dot K$ given in (\ref{sskeom}) can be cast into the form
\bea
K' &=& -\fft{4r}{\Delta}\, 
\fft{(\tau' \, \bar\tau' -\dot\tau \dot{\bar\tau})}{(\tau-\bar\tau)^2}
+ \fft{r}{4\Delta}\, (g'^2-\dot g^2)\ ,\nn\\
\dot K &=& -\fft{4r}{\Delta}\, 
\fft{(\tau'\, \dot{\bar\tau} + \dot\tau\, {\bar\tau}')}{(\tau-\bar\tau)^2}
+ \fft{r}{2\Delta}\, \dot g g'\ .\label{kexp}
\eea
(After writing out these expressions in terms of $\chi$ and $f$, one should 
again make the replacement $\chi=-\im \gamma$.)
This shows that $K$ is invariant under the $SL(2,\R)$ transformation.  In 
fact, it is also invariant under the entire $O(2,1)/J$ symmetry.  In 
particular, this explains why the expressions for $K'$ and $\dot K$ that we 
obtained in (\ref{keq}) for the charged solutions are the same as they would 
be for uncharged solutions.

     By varying the Lagrangian (\ref{fgamlag}) with respect to local 
infinitesimal $SL(2,\R)$ transformations, we can read off the Noether 
currents that generate the symmetry.  They are proportional to
\bea
J_1 &=& r\, e^{-f} \, \nabla \gamma\ ,\nn\\
J_2 &=& r\, \nabla f - \ft12 \Delta\, r\, e^{-f}\, \gamma\nabla\gamma\ ,
\label{noether}\\
J_3 &=& r\, \nabla\gamma - r\, \gamma\, \nabla f - \ft14\, \Delta \, r\, 
e^{-f}\, \gamma^2\, \nabla\gamma\ .
\eea
The associated conserved charges, which are independent of $r$, are obtained
by integrating the $r$-components of these currents over all $z$.  In 
particular, the first current in (\ref{noether}) integrates to give the 
electric charge carried by the field strength $F$:
\bea
Q &=& \fft{1}{4\pi}\, \int *F\, e^{-a\phi} = \fft{1}{4\pi}\, 
\int *(d\gamma\wedge dt) e^{-a\phi} \ ,\nn\\
&=& \int r\, e^{-f}\, \gamma'\, dz\ .\label{charge}
\eea

     The multi-scalar equations (\ref{mseom}) can be analysed in a similar 
manner.  In this case, we find that the equations for $\Phi_\a$ and 
$\gamma_\a$ can be obtained from the Lagrangian
\be
L= \sum_{\a=1}^N\Big((\nabla\Phi_\a)^2 -4 e^{-\Phi_\a}\, 
(\nabla\gamma_\a)^2 \Big) \ .\label{mslag}
\ee
Clearly this has an $(O(2,1)/J)^N$ symmetry, which again may be used to
obtain new solutions from old ones, in the same way as we described for the
single-scalar case above.  In particular, we could begin with a pure 
Einstein-Dilaton solution with no electric charges at all, and use the $N$ 
independent $O(2,1)/J$ symmetries to turn on the $N$ charges.  For example, 
by starting from the uncharged solution where $Y$ and $\Phi_\a$ are 
independent harmonic functions, and the functions $\gamma_\a$ are 
appropriately chosen constants, we can reproduce the general class of 
charged solutions given in (\ref{eom}).

\subsection{Dyonic solutions}

   Now let us consider the situation when the 2-form field strength carries 
both electric and magnetic charge, in which case $F$ takes the form
\be
F=(\gamma'\, dr + \dot\gamma\, dz)\wedge dt + r\, e^{a\phi-2U}\, 
(\tilde\gamma' \, dz -\dot{\tilde \gamma}\, dr)\wedge d\theta\ ,
\ee
and the equations of motion become
\bea
&&\nabla^2 U = \ft14 e^{-a\phi-2U}\, (\nabla \gamma)^2 + \ft14 
e^{a\phi-2U}\, (\nabla \tilde\gamma)^2 \ ,\nn\\
&&\nabla^2 \phi = \ft12 a\, e^{-a\phi-2U}\, (\nabla \gamma)^2 -\ft12 a\,
e^{a\phi-2U}\, (\nabla \tilde\gamma)^2 \ ,\nn\\
&&\nabla\cdot (e^{-a\phi-2U}\, \nabla \gamma)= 0 \ ,\label{dyoneom}\\
&&\nabla\cdot (e^{a\phi-2U}\, \nabla \tilde\gamma) = 0 \ ,\nn\\
\eea
together with the equations that determine $K$ in (\ref{metric}), namely
\bea
K'&=& \ft14 r\, e^{-a\phi-2U}\,(\dot\gamma^2-{\gamma'}^2)+
\ft14 r\, e^{a\phi-2U}\,(\dot{\tilde\gamma}^2-{\tilde\gamma}'^2) +r\, (U'^2 -
\dot U^2) +\ft14 r\, (\phi'^2 -\dot \phi^2)\ ,\nn\\
\dot K &=& -\ft12 r\, e^{-a\phi-2U}\, \dot\gamma\, \gamma'
-\ft12 r\, e^{a\phi-2U}\, \dot{\tilde\gamma}\, {\tilde\gamma}' + 2 r\, \dot U\, 
U' +\ft12 r\, \dot\phi\, \phi' \ .\label{dyonkeom}
\eea

     Defining two new functions $q_1$ and $q_2$ in terms of $\phi$ and 
$U$, given by
\be
a\, q_1 = \phi + 2 a \, U\ ,\qquad a\, q_2= -\phi + 2a\, U\ ,
\ee
the equations of motion become
\bea
&&\nabla^2 q_1 = e^{-\a q_1 -(1-\a) q_2}\, (\nabla \gamma)^2\ ,\nn\\
&&\nabla^2 q_2 = e^{-\a q_2 -(1-\a) q_1}\, (\nabla \tilde\gamma)^2\ ,\nn\\
&&\nabla\cdot(e^{-\a q_1 -(1-\a)q_2}\, \nabla \gamma) = 0\ ,\label{qeq}\\
&&\nabla\cdot(e^{-\a q_2 -(1-\a)q_1}\, \nabla \tilde\gamma) = 0\ ,
\eea
where the constant $\a$ is given by $\a=\ft12(1+a^2)=\ft12\Delta$.  For 
general values of $a$ it seems not to be possible to find solutions to these 
equations other than the rather trivial case where $q_1=q_2$ and 
$\gamma=\tilde\gamma$, for which the equations reduce to
\be
\nabla^2 q_1= e^{-q_1}\, (\nabla\gamma)^2\ , \qquad \nabla\cdot(e^{-q_1}\, 
\nabla\gamma)=0\ .
\ee
This system can be solved by analogous methods to those that we used in 
section 5.1, by 
writing $q_1$ and $\gamma$ as functions of a single harmonic function 
$\psi$.  This reduces it to the Liouville equation.  Note that in this 
special solution, the dilaton $\phi$ is zero.

     There are two special values of $a$ for which more general solutions 
can be given.  Firstly, if we take $a=1$, which implies that $\a=1$, we see 
that the equations (\ref{qeq}) decouple.  They can be solved by taking $q_1$ 
and $\gamma$ to be functions of an harmonic function $\psi_1$, and $q_2$ and 
$\tilde\gamma$ to be functions of an independent harmonic function $\psi_2$, 
reducing the system to two independent Liouville equations, with the 
solutions
\bea
&& e^{-\ft12 q_1} =e^{-\psi_1} -c_1^2\, e^{\psi_1}\ ,\qquad
\gamma=\sqrt2 c_1\, e^{2\psi_1}(1-c_1^2\, e^{2\psi_1})^{-1}\ ,\nn\\
&& e^{-\ft12 q_2} =e^{-\psi_2} -c_2^2\, e^{\psi_2}\ ,\qquad
\tilde\gamma=\sqrt2 c_2\, e^{2\psi_2}(1-c_2^2\, e^{2\psi_2})^{-1}\ ,
\eea
where $c_1$ and $c_2$ are arbitrary constants.  The other special case is 
$a=\sqrt3$, which implies $\a=2$.  Now, if we take $q_1$, $q_2$, $\gamma$ 
and $\tilde\gamma$ all to be functions of a single harmonic function 
$\psi$, the equations reduce to the $SL(3,R)$ Toda equations
\be
\fft{d^2 q_1}{d\psi^2}=e^{2q_1-q_2}\ ,\qquad  
\fft{d^2 q_2}{d\psi^2}=e^{2q_2-q_1}\ ,\label{toda}
\ee
where we have absorbed two constants of integration into rescalings of 
$\gamma$ and $\tilde\gamma$ together with shift transformations of $q_1$ and 
$q_2$.  The solutions for $\gamma$ and $\tilde \gamma$ are 
$\gamma=dq_1/d\psi$ and $\tilde\gamma=dq_2/d\psi$.  The general solution to 
the Toda equations (\ref{toda}) can be written as
\bea
e^{-q_1}&=& \fft{c_1\, e^{\mu_1\psi}}{\nu_1(\nu_1-\nu_2)} +
            \fft{c_2\, e^{\mu_2\psi}}{\nu_2(\nu_1-\nu_2)} +
            \fft{e^{-(\mu_1+\mu_2)\psi}}{c_1 c_2\nu_1\nu_2}\ ,\nn\\
e^{-q_2}&=& \fft{e^{-\mu_1\psi}}{c_1\nu_1(\nu_1-\nu_2)} +
            \fft{e^{-\mu_2\psi}}{c_2\nu_2(\nu_1-\nu_2)} +
            \fft{c_1c_2\, e^{(\mu_1+\mu_2)\psi}}{\nu_1\nu_2}\ ,
\eea
where $\mu_1$, $\mu_2$, $c_1$ and $c_2$ are arbitrary constants,
$\nu_1\equiv 2\mu_1+\mu_2$ and $\nu_2\equiv 2\mu_2+\mu_1$.  This case where 
$a=\sqrt3$ is actually realised in maximal supergravity in $D=4$.

     We can examine the symmetries of the dyonic system of equations in a 
manner analogous to that discussed in section 5.2 for the purely electric 
case.  It turns out that for arbitrary $a$ the system possesses only the 
following symmetries:
\bea
&&\gamma\rightarrow e^\lambda\, \gamma\ ,\qquad \tilde\gamma\rightarrow 
e^{\tilde\lambda}\, \tilde\gamma\ ,\nn\\
&&\phi\rightarrow \phi +\ft12 (\lambda-\tilde\lambda)\ ,\qquad U\rightarrow 
U +\ft14 (\lambda+\tilde\lambda)\ ,
\eea
where $\lambda$ and $\tilde\lambda$ are constants.  However, when $a=1$ (and 
for no other values), the symmetry is enlarged to $(O(2,1)/J)^2$.  
This is not surprising, since at this value of $a$ we saw that the system 
decouples into two independent Liouville equations, which, as we saw 
previously, will each have an $O(2,1)/J$ symmetry.  In this $a=1$ case the 
symmetry can be used to produce new solutions from old ones.

\section{Conclusions}

     In this paper, we have studied classes of non-extremal black hole
solutions in 4-dimensional supergravity.  Owing to special properties of
four dimensions, it turns out that solutions can be given, for 
arbitrary distributions of black holes along an axis, in terms of harmonic 
functions.  The forces between the individual black holes are balanced by 
``struts'' located between them, on which the curvature has conical 
singularities.  These struts, and their associated singularities, disappear 
in the cases of principal interest to us, namely for simply-periodic arrays 
where there is a net balance of forces on each black hole.  In an appropriate 
dense limit, such a periodic array can be used to describe a vertical 
dimensional reduction to black hole solutions in three dimensions.

     The cases that we were able to solve included multi-charge black holes 
in four dimensions, where several different field strengths and dilatonic 
scalars in the four-dimensional theory are involved.  These solutions can 
also be diagonally oxidised in a straightforward fashion, to give 
multi-charge non-extremal $(D-4)$-branes in $D\ge5$ dimensions.  
One can also investigate the oxidation of such solutions all the way back to 
the original $D=11$ supergravity from which the four-dimensional theory can 
be viewed as originating.

     We also 
examined the special case of the dyonic black hole in four dimensions.  
This involves a single field strength, carrying both electric and magnetic 
charges.   We were able to construct a particular class of multi-centre 
solutions, describing arbitrary distributions of dyonic black holes along an 
axis, where each individual black hole carries the same ratio of electric 
to magnetic charge. This is sufficient for the purposes of performing a 
vertical reduction to $D=3$.  It would also be very interesting if one could 
obtain more general solutions in which the individual electric and magnetic 
constituent charges could be separated in different centres.  This ``pulling 
apart'' of the dyon would enable its stability properties to be investigated 
in detail.  This would be of great interest because even in the extremal 
limit, the dyonic black hole has non-vanishing binding energy; it is 
negative, implying that there should be a tendency for the electric and 
magnetic charges to separate into individual electric and magnetic black 
holes \cite{gk}.  Unfortunately, we were unable to solve the equations in this 
case, although it may nevertheless be that they would be tractable given 
suitable techniques for solving them.

\end{document}